\documentclass[preprintnumbers,prd,onecolumn,floatfix,superscriptaddress, nofootinbib]{revtex4-2}
\usepackage{array}
\usepackage{graphicx}
\usepackage{setspace}
\usepackage{subfig}
\usepackage{epsfig}
\usepackage{bm}
\usepackage{amssymb}
\usepackage{float}
\usepackage{amsmath}
\usepackage{dcolumn}
\usepackage[colorlinks]{hyperref}
\usepackage[usenames,dvipsnames]{color}
\usepackage{enumitem}
\usepackage{orcidlink} 

\hypersetup{ breaklinks=true, pdfstartview={FitH}, colorlinks=true, linkcolor=blue, citecolor=red, filecolor=magenta, urlcolor=blue, anchorcolor=green, linktocpage=true }

\begin{document}

\title{Study of the dark energy models in $f(Q, \mathcal{L}_m)$ gravity in different formalisms}
\author{Vidhi Tayal\orcidlink{0009-0003-4581-8850}}
\email{vidhitayal2701@gmail.com}
\affiliation{School of Engineering and Applied Sciences, Bennett University, Greater Noida, India}
\author{Shaily\orcidlink{0000-0002-9598-5900}}
\email{shailytyagi.iitkgp@gmail.com}
\affiliation{School of Engineering and Applied Sciences, Bennett University, Greater Noida, India}
\author{Vimal Kumar\orcidlink{0000-0003-1038-1102}}
\email{Vimal.Kumar@bennett.edu.in}
\affiliation{School of Computer Science Engineering and Technology, Bennett University, Greater Noida, India}
\author{J. K. Singh\orcidlink{0009-0000-6037-6702}}
\email{jksingh@nsut.ac.in}
\affiliation{Department of Mathematics, Netaji Subhas University of Technology, New Delhi-110078, India}
\author{Anirudh Pradhan\orcidlink{0000-0002-1932-8431}}
\email{pradhan.anirudh@gmail.com}
\affiliation{Centre for Cosmology, Astrophysics and Space Science (CCASS), GLA University, Mathura-281 406, Uttar Pradesh, India}

\begin{abstract}
We study the late-time cosmological evolution of the Universe in the framework of a flat Friedmann-Lemaˆıtre-Robertson-Walker spacetime in  $f(Q,\mathcal{L}_m)$  gravity, where the model is $ f(Q, \mathcal{L}_m)=-\alpha Q+\lambda (2\mathcal{L}_m )^2+\beta $. The cosmic dynamics of the Friedmann equations are discussed using various equation-of-state (EoS) parameters like Chevallier–Polarski–Linder (CPL), Jassal–Bagla–Padmanabhan (JBP), Barboza–Alcaniz (BA), and Logarithmic (LOG). We use joint observational datasets of Hubble (OHD), DES five-year Type Ia supernovae (DES~5YR), and the Baryon Acoustic Oscillation (BAO) measurements to obtain the best-fit values of model parameters, which show good agreement with current observations. These models exhibit a smooth transition from early deceleration to late-time acceleration. The equation of state parameter displays quintessence-like behaviour over a wide redshift range and gradually converges its behaviour from quintessence-like to phantom-like behaviour as $ z\rightarrow-1 $, except in the BA case. Furthermore, we investigate the physical characteristics of the model in order to assess its physical viability. We also analyse the energy conditions and examine the stability of the proposed model.
\end{abstract}

\maketitle

\section{Introduction}\label{intro}

\qquad Modern observations such as Type Ia supernovae \cite{SupernovaSearchTeam:1998fmf, SupernovaCosmologyProject:1998vns}, the cosmic microwave background \cite{Planck:2018vyg}, and galaxy surveys show that the Universe is expanding faster. It also indicates that the Universe is experiencing accelerated expansion \cite{Koivisto:2005mm, SDSS:2005xqv, Daniel:2008et}. The usual theory of gravity, General Relativity (GR), can explain many things, but it needs a mysterious dark energy to explain this accelerated expansion. The standard cosmological framework, known as the $\Lambda$-Cold Dark Matter ($\Lambda $CDM) model, treats dark energy as a constant component that does not vary with space or time. Although $\Lambda $CDM works well in many cases, there are still some problems, such as the disagreement between different measurements of the Hubble constant and the observed value of  $\Lambda $ is vastly smaller than the value of  $\Lambda $ predicted by quantum physics. Because of these issues, many researchers explore modified theories of gravity. 

These issues motivate researchers to explore alternative approaches, including modified gravity and geometry–matter coupling theories. One of the earliest and most influential frameworks is $f(R)$ gravity, studied widely by Sotiriou and Faraoni (2010) \cite{Sotiriou:2008rp} and De Felice and Tsujikawa (2010) \cite{DeFelice:2010aj}, where the Ricci scalar $R$ is replaced by a function $f(R)$ \cite{Dunsby:2010wg, Molano:2025fyj}. This approach can explain cosmic acceleration but often leads to higher-order and sometimes unstable field equations. Another important class is $f(T)$ gravity, developed by Bengochea and Ferraro ($2009$) \cite{Bengochea:2008gz} and later extended by Cai et al. \cite{Cai:2015emx}, where gravity is described using torsion instead of curvature; it provides second-order equations but faces challenges with local Lorentz invariance. An alternative approach that has gained considerable attention in recent years is $f(Q)$ gravity, which is developed within the setting of symmetric teleparallel geometry. In this description, gravity is not associated with curvature or torsion but instead arises from the non-metricity scalar $Q$. Comprehensive discussions of this theory can be found in the recent review by Heisenberg ($2024$) \cite{Dowling:2024wvo} and others \cite{Koussour:2022nsc, Khyllep:2021pcu, Koussour:2022sab, Myrzakulov:2022akb, Koussour:2023rly}. This theory naturally yields second-order equations and can reproduce late-time acceleration without explicitly reducing a cosmological constant. Several extensions of modified gravity theories have also been proposed by incorporating direct couplings between matter and geometry. A well-known example is $f(R, T)$ gravity introduced by Harko et al. (2011) \cite{Harko:2011kv} and later extended by others \cite{Singh:2023gxd, Singh:2018xjv}. Similarly, $f(Q, T)$ gravity developed by Xu et al. (2019) \cite{Xu:2019sbp, Xu:2020yeg, ElBourakadi:2023ufz, Singh:2024ckh, Singh:2024kez, Nagpal:2018mpv} extends the non-metricity framework by allowing the gravitational action to depend on both $Q$ and the trace of the energy–momentum tensor $T$. These theories generally lead to modified conservation relations and can generate a wider variety of cosmological dynamics. Additional modifications like $f(T, B)$ gravity, explored by Bahamonde et al.  \cite{Bahamonde:2021srr}, combine torsion and boundary terms in the teleparallel formulation. Likewise, Gauss–Bonnet models like $f(G)$ gravity, studied by Nojiri and Odintsov \cite{Nojiri:2017ncd}, add extra geometric terms to the equations of gravity \cite{Bamba:2017cjr, Singh:2023bjx, DeFelice:2008wz, Singh:2024aml, Singh:2022gln}. Such modifications can lead to new cosmological phenomena and provide alternative explanations for cosmic acceleration. Despite these developments, many modified gravity models face certain challenges. While they are often capable of explaining the accelerated expansion of the Universe, some of them encounter issues related to theoretical consistency, stability conditions, or compatibility with observational constraints.

Among the different extensions of gravitational theory, $f(Q, \mathcal{L}_m)$ gravity has gained considerable attention in recent years as a useful framework for cosmological investigations. In this formulation, the gravitational action is constructed from the non-metricity scalar $Q$ together with the matter Lagrangian $\mathcal{L}_m$, thereby linking the matter sector directly with the geometric sector. This matter-geometry coupling modifies the resulting cosmological dynamics and allows the theory to account for the accelerated expansion of the Universe without explicitly introducing a cosmological constant. Since the matter Lagrangian contributes directly to the gravitational action, the resulting dynamics may also depend on the properties of the matter sector. In addition, the theory naturally gives rise to modified conservation equations, which may affect the dynamical evolution of the Universe. A further notable aspect of $f(Q, \mathcal{L}_m)$ gravity is that certain models can recover the background expansion behaviour of the standard $\Lambda$CDM scenario while still allowing differences in the growth of cosmic structures. Such departures may serve as useful observational signatures for distinguishing these models from the standard cosmological framework. These features, together with the increasing number of studies reported in the literature, have made $f(Q, \mathcal{L}_m)$ gravity a flexible setting for examining cosmological dynamics and the large-scale evolution of the Universe. Consequently, interest in this framework has grown rapidly in recent years, with contributions to both its theoretical formulation and its cosmological applications. 

Recent investigations of the $f(Q, \mathcal{L}_m)$ framework \cite{Myrzakulov:2024esv, Singh:2026wsg, Hazarika:2024alm, Myrzakulov:2025yxy} have also focused on particular functional forms that admit analytical cosmological solutions. One such model is  $f(Q, \mathcal{L}_m)=-\alpha Q+\lambda (2\mathcal{L}_m )+\beta$, which admits explicit solutions characterizing the late-time cosmological evolution of the Universe. Observational analyses based on $H(z)$ Pantheon+SH0ES, and BAO datasets show that this model can describe the transition of the Universe from a decelerating phase to the present accelerating phase. At the same time, its background evolution remains close to that predicted by the standard $\Lambda $CDM. Another study looked at a model of the form $f(Q, \mathcal{L}_m)=\alpha Q+\beta (\mathcal{L}_m)^n$. In this case, the cosmological dynamics depend strongly on the value of the parameter $n$. They found that the universe accelerates when $n>2$, while $n=1$ behaves like a universe filled with non-relativistic matter. In another work, Myrzakulov et al. \cite{Myrzakulov:2024esv} investigated a viscous cosmological scenario within $f(Q, \mathcal{L}_m)$ gravity using the linear form $f(Q, \mathcal{L}_m)=\alpha Q+\beta \mathcal{L}_m$. Their analysis highlighted the role of bulk viscosity in modifying the expansion history of the Universe and provided constraints on the model parameters. 

Further progress was made by Myrzakulov et al. \cite{Myrzakulov:2025yxy}, who examined the validity of standard energy conditions NEC, WEC, SEC and DEC within $f(Q, \mathcal{L}_m)$ gravity and identified parameter ranges where the theory remains physically viable. Additional related studies on $f(Q)$ and $f(Q, T)$ gravity by various authors between 2019 and 2024 \cite{BeltranJimenez:2019tme, Koussour:2023rly} have contributed significantly to the understanding of non-metricity-based gravitational theories. These works explored aspects such as cosmological solutions, properties of non-metricity, and perturbation behaviour, thereby providing an important theoretical foundation for the more general  $f(Q, \mathcal{L}_m)$ framework. Collectively, these works indicate that $f(Q, \mathcal{L}_m)$ gravity offers a versatile theoretical setting capable of explaining late–time cosmic acceleration, modifying the effective gravitational coupling, and generating observational signatures that differ from those predicted by standard cosmology.

Alongside modified gravity theories, equation-of-state (EoS) parametrisations provide a useful phenomenological framework for examining the late-time expansion of the Universe. Instead of restricting the EoS parameter to a constant value, these parametrisations permit $w(z)$ to evolve with redshift, allowing a more flexible representation of the dynamical nature of dark energy. Such approaches are particularly useful for comparing theoretical predictions with observational data without imposing a specific underlying model.
 Several forms of EoS parametrisations have been widely employed in cosmological studies, including the Chevallier Polarski Linder (CPL), Jassal Bagla Padmanabhan (JBP), Barboza-Alcaniz (BA), and logarithmic (LOG) parametrisations. These parametrisations have been extensively employed in cosmological analyses using observational datasets such as Hubble parameter measurements, Type Ia supernovae, and baryon acoustic oscillations. Recent investigations have further explored the role of evolving EoS parametrisations in testing dynamical dark energy models and deviations from the standard $\Lambda CDM$. Several recent studies have explored the role of dynamical EoS parametrisations within modified gravity and alternative cosmological frameworks. For instance, CPL, JBP, and BA parametrisations have been investigated in minimally modified gravity and Gauss–Bonnet gravity models, as well as in the context of $f(Q)$ gravity and observational cosmology\cite{Wolf:2025jlc, Arora:2025msq, Mukherjee:2025yrp, Mazumdar:2025smh}

Motivated by these developments, in the present work we investigate the late-time cosmological evolution of the Universe within the framework of $f(Q, \mathcal{L}_m)$ gravity by considering the model $f(Q, \mathcal{L}_m)=-\alpha Q+\lambda (2\mathcal{L}_m )^2+\beta$, which introduces a nonlinear matter-geometry coupling. This model is proposed by Hazarika et al. \cite{Hazarika:2024alm}. When studied in the context of a flat FLRW universe, it is able to explain the observed accelerated expansion, offering an alternative to the usual dark-energy explanation. This model can naturally produce a wide variety of expansion histories for the Universe without requiring extra modifications to the geometric part of gravity. Our main goal is to rebuild the late-time cosmic evolution using several commonly used dark-energy parametrisations, including CPL, BA, JBP, and LOG forms. After obtaining these reconstructed evolutions, we compare them with current astronomical observations, such as the Hubble dataset, DES 5YR dataset and the BAO dataset, to check how well the model matches real cosmic behaviour.

\qquad The content of the work is organised as follows: In section \ref{framework}, we review the field equations of $ f(Q, \mathcal{L}_m) $ gravity with a specific functional form of the model $f(Q, \mathcal{L}_m)$. In Section \ref{outline}, we introduce the various dynamical EoS parametrisations to solve the Friedmann equations and derive the corresponding cosmological parameters for each model. In Section \ref{data}, we present different observational datasets that are employed to constrain the model parameters. We discuss the reconstructed cosmological parameters and analyse their evolutionary behaviour within the proposed framework in Section \ref{param}. The validity of the energy conditions and the stability of the model are examined in Sections \ref{ec} and \ref{sa}, respectively. Finally, we conclude our results and their cosmological implications in Section \ref{conclusion}.

\section{ The $f(Q,\mathcal{L}_m)$ gravity}\label{framework}
The $f(Q, \mathcal{L}_m)$ class of theories generalises symmetric teleparallel gravity by allowing the action to depend on both the non-metricity scalar $Q$ and the matter Lagrangian density $\mathcal{L}_m$ \cite{Harko:2018gxr, Myrzakulov:2024esv, Harko:2014gwa}. The gravitational action is expressed as described in \cite{Hazarika:2024alm}
\begin{equation}\label{1}
A = \int f(Q,\mathcal{L}_m)\sqrt{-g}\, d^4x,
\end{equation}
where $g$ denotes the determinant of the spacetime metric. Within this framework, the gravitational dynamics are governed by the non-metricity of the connection, whereas the dependence on $\mathcal{L}_m$ gives rise to a direct interaction between matter and geometry. 

The non-metricity tensor is defined as
\begin{equation}\label{2}
Q_{\gamma\mu\nu} = -\nabla_\gamma g_{\mu\nu},
\end{equation}
and measures the failure of the metric to be preserved under parallel transport \cite{BeltranJimenez:2017tkd, Lazkoz:2019sjl}. Two traces of this tensor are given by
\begin{equation}\label{3}
Q_\beta = g^{\mu\nu}Q_{\beta\mu\nu}, \qquad 
\tilde{Q}_\beta = g^{\mu\nu}Q_{\mu\beta\nu}.
\end{equation}
A key geometric quantity in symmetric teleparallel gravity is the disformation tensor,
\begin{equation}\label{4}
L^\beta{}_{\alpha\gamma}
= \frac12 g^{\beta\eta}
\left(Q_{\gamma\alpha\eta}+Q_{\alpha\eta\gamma}-Q_{\eta\alpha\gamma}\right),
\end{equation}
which characterises how the connection deviates from the Levi Civita connection in a torsion-free and curvature-free geometry. Using this, the superpotential is defined as
\begin{equation}\label{5}
P^\beta{}_{\mu\nu}
= -\frac12 L^\beta{}_{\mu\nu}
+ \frac14 (Q^\beta-\tilde{Q}^\beta)\, g_{\mu\nu}
- \frac14 \delta^\beta_{(\mu} Q_{\nu)},
\end{equation}
and the associated non-metricity scalar is given as
\begin{equation}\label{6}
Q = - Q_{\beta\mu\nu} P^{\beta\mu\nu}.
\end{equation}

Variation of the action with respect to the metric yields the field equations \cite{Harko:2018gxr, Myrzakulov:2025yxy}, 
\textbf{\begin{equation}\label{7}
\frac{2}{\sqrt{-g}} \nabla_{\alpha} \left( \sqrt{-g}\, f_Q P^{\alpha}{}_{\mu\nu} \right)
+ f_Q \left( P_{\mu\alpha\beta} Q^{\alpha\beta}{}_{\nu}
- 2 Q^{\alpha\beta}{}_\mu P_{\alpha\beta\nu} \right)
+ \frac{1}{2} f g_{\mu\nu}
= \frac{1}{2} f_{\mathcal{L}_m} \left( g_{\mu\nu} \mathcal{L}_m - T_{\mu\nu} \right),
\end{equation}}
where the contributions from the geometry and matter sectors enter through the derivatives 

$f_Q = \frac{\partial f}{\partial Q}, \qquad$
$f_{\mathcal{L}_m} = \frac{\partial f}{\partial \mathcal{L}_m}$.

The matter energy-momentum tensor is defined as  
\begin{equation}\label{8}
T_{\mu\nu} = -\frac{2}{\sqrt{-g}}
\frac{\delta(\sqrt{-g}\mathcal{L}_m)}{\delta g^{\mu\nu}}
= g_{\mu\nu}\mathcal{L}_m - 2\frac{\partial \mathcal{L}_m}{\partial g^{\mu\nu}}
\end{equation}
and variation with respect to the independent connection introduces the hypermomentum density, which encodes the response of matter fields to changes in the connection \cite{Harko:2018gxr}.

A direct consequence of the non-minimal dependence on $\mathcal{L}_m$ is that the energy-momentum tensor is generally not conserved. The divergence $\nabla_\mu T^\mu{}_\nu$ contains additional terms depending on $Q$, $\mathcal{L}_m$, and derivatives of $f(Q, \mathcal{L}_m)$, indicating that test particles are subject to an extra force and do not follow geodesics.

For cosmological applications, we adopt the spatially flat Friedmann-Lemaître-Robertson-Walker (FLRW) metric,
\begin{equation}\label{9}
ds^2 = -dt^2 + a^2(t)(dx^2 + dy^2 + dz^2),
\end{equation}
for which the non-metricity scalar becomes $Q = 6H^2$, where $H=\dot{a}/a$ is the Hubble parameter. Assuming a perfect-fluid matter source with energy density $\rho$ and pressure $p$, we get
\begin{equation}\label{10}
T_{\mu\nu} = (\rho+p)u_\mu u_\nu + p g_{\mu\nu}.
\end{equation}
For the above metric the Friedmann equations are obtained as \cite{Hazarika:2024alm,Myrzakulov:2024esv,Myrzakulov:2025yxy,Myrzakulov:2024kuu}
\begin{equation}\label{11}
3H^2 = \frac{1}{4f_Q}\left[f - f_{\mathcal{L}_m}(\rho + \mathcal{L}_m)\right],
\end{equation}
\begin{equation}\label{12}
\dot{H} + 3H^2 + \frac{\dot{f}_Q}{f_Q}H 
= \frac{1}{4f_Q}\left[f + f_{\mathcal{L}_m}(p - \mathcal{L}_m)\right].
\end{equation}

To examine the effects of matter geometry coupling, we consider the model \cite{Hazarika:2024alm}
\begin{equation}\label{13}
f(Q,\mathcal{L}_m)= -\alpha Q + \lambda (2\mathcal{L}_m)^2 + \beta,
\end{equation}
where $\alpha$, $\beta$, and $\lambda$ are constants, and we take $\mathcal{L}_m=\rho$. In cosmological applications, the matter Lagrangian for a perfect fluid is often chosen as $\mathcal{L}_m=\rho$, which simplifies the coupling between matter and geometry and has been widely adopted in modified gravity theories \cite{Harko:2011kv, Xu:2019sbp}.
In this case,
\begin{equation}\label{14}
f_Q = -\alpha, \qquad f_{\mathcal{L}_m} = 8\lambda\rho.
\end{equation}
Now, the above derivatives, Eq. (\ref{11}), become
\begin{equation}\label{15}
3H^{2}=\frac{1}{4(-\alpha)}\left[-6\alpha H^{2} + 4\lambda\rho^{2} + \beta 
 - 16\lambda\rho^{2}\right].
\end{equation}
Simplifying the expression yields
\begin{equation}\label{16}
\rho=\frac{\sqrt{6\alpha H^{2}+\beta}}{2\sqrt{3\lambda}}.
\end{equation}

Now substituting $f_{Q}=-\alpha$, the expression for $f$, and $f_{L_{m}}=8\lambda\rho$, 
the second Friedmann equation (\ref{12}) becomes
\begin{equation}\label{17}
\alpha\dot{H}+\frac{3\alpha}{2}H^{2}
= \lambda\rho^{2} - \frac{\beta}{4} - 2\lambda \rho p.
\end{equation}
Solving for $p$ gives
\begin{equation}\label{18}
p=-\frac{6\alpha(\dot{H}+H^{2})+\beta}
{2\sqrt{3\lambda}\,\sqrt{6\alpha H^{2}+\beta}}.
\end{equation}
The above results reveal that in this nonlinear matter-geometry coupling model, both the effective energy density and pressure acquire explicit dependence on the coupling parameters and on the Hubble expansion rate. The square-root structure in the equations allows the model to reproduce accelerated expansion for appropriate choices of $(\alpha,\lambda,\beta)$, making this framework a viable candidate for describing late-time cosmological dynamics within symmetric teleparallel gravity.

\section{EoS Parametrizations and Dynamical Analysis}\label{outline}
To examine the dynamical implications of the nonlinear matter–geometry coupling, we compute the redshift evolution of the Hubble parameter and related cosmological quantities for four widely used EoS parametrisations: Chevallier–Polarski–Linder (CPL), Barboza–Alcaniz (BA), Jassal–Bagla–Padmanabhan (JBP) and Logarithmic (LOG). These models are designed to check whether dark energy changes over time, instead of staying constant like in the standard cosmological constant. Each model uses two main parameters. The first parameter $w_0$ specifies the current value of the equation-of-state parameter. The second parameter $w_1$ shows how this value changes as the universe expands. Using these two parameters, we study both constant and changing behaviours of dark energy within our modified gravity framework.

\subsection{Chevallier–Polarski–Linder (CPL) Parametrisation}
 The CPL form provides a first-order Taylor expansion of $w(a)$ around the present scale factor $a_0=1$ \cite{Chevallier:2000qy}, which is given as
\begin{equation}\label{19}
w_{CPL}(a) = w_0 + w_1(1-a), 
\qquad 
a = \frac{1}{1+z}.
\end{equation}

Eqs. (\ref{16}), (\ref{18}), (\ref{19}) and equation of state $p = w \rho$  gives
\begin{equation}\label{20}
H_{CPL}(z)=\sqrt{
\frac{
\left(6\alpha H_{0}^{2}+\beta\right)
(1+z)^{2\left(1+w_{0}+w_{1}\right)}
\exp\!\left(-\frac{2w_{1}z}{1+z}\right)
-\beta
}{6\alpha}
}.
\end{equation}

Using the expression above, the energy density and pressure for CPL are
reconstructed as
\begin{equation}\label{21}
\rho_{CPL}(z) =
\frac{
\sqrt{
\exp\!\left(-\frac{2w_1 z}{1+z}\right)
(1+z)^{2(1+w_0+w_1)}
\left(6\alpha H_0^2 + \beta\right)
}
}{
2\sqrt{3}\,\sqrt{\lambda}
},
\end{equation},
\begin{equation}\label{22}
p_{CPL}(z) =
\frac{
\sqrt{
\exp\!\left(-\frac{2w_1 z}{1+z}\right)
(1+z)^{2(1+w_0+w_1)}
\left(6\alpha H_0^2 + \beta\right)
}
\left(w_0 + zw_0 + zw_1\right)
}{
2\sqrt{3}\,(1+z)\,\sqrt{\lambda}
}.
\end{equation}

For the CPL parametrisation, the deceleration parameter can be expressed as

\begin{equation}\label{23}
q_{CPL}(z) = -1+\frac{(1+z)}{H}\frac{dH}{dz}=-1 +
\frac{
(1+z)^{1+2w_0+2w_1}
\left(6\alpha H_0^2 + \beta\right)
\left[1+w_0+z(1+w_0+w_1)\right]
}{
-\beta \exp\!\left(-\frac{2w_1 z}{1+z}\right)
+
(1+z)^{2(1+w_0+w_1)}
\left(6\alpha H_0^2 + \beta\right)
},
\end{equation}

\begin{equation}\label{24}
\begin{aligned}
j_{CPL}(z)&=q(z)\,[2q(z)+1]+(1+z)\frac{dq(z)}{dz} \\ &
=\frac{
- e^{\frac{2 z w_1}{1+z}} \beta
+ (1+z)^{2(w_0+w_1)} (6 H_0^2 \alpha + \beta)
\Big[
(1+z)^2 w_0 (1+2w_0)
+ (1+z)(1+z+4 z w_0)w_1
+ 2 z^2 w_1^2
\Big]
}{
- e^{\frac{2 z w_1}{1+z}} \beta
+ (1+z)^{2(1+w_0+w_1)} (6 H_0^2 \alpha + \beta)
}.
\end{aligned}
\end{equation}

\subsection{Barboza–Alcaniz (BA) Parametrisation}
The BA parametrisation introduces a nonlinear correction designed to maintain regularity across both early and late cosmic times \cite{Barboza:2008rh}. It is defined by
\begin{equation}\label{25}
w_{BA}(a)=w_0+w_1\frac{1-a}{a^2+(1-a)^2},  
\qquad 
a = \frac{1}{1+z}.
\end{equation}
With this EoS parameter, one can obtain
\begin{equation}\label{26}
H_{BA}(z)=\sqrt{
\frac{
\left(6\alpha H_{0}^{2}+\beta\right)
(1+z)^{2\left(1+w_{0}\right)}
\left(1+z^{2}\right)^{w_{1}}
-\beta
}{6\alpha}.
}
\end{equation}
From this, the corresponding cosmological parameters are reconstructed as:
\begin{equation}\label{27}
\rho_{BA}(z) =
\frac{
\sqrt{
(1+z)^{2(1+w_0)}
(1+z^2)^{w_1}
\left(6\alpha H_0^2 + \beta\right)
}
}{
2\sqrt{3}\,\sqrt{\lambda}
},
\end{equation}

\begin{equation}\label{28}
p_{BA}(z) =
\frac{
\sqrt{
(1+z)^{2+2w_0}
(1+z^2)^{w_1}
\left(6\alpha H_0^2 + \beta\right)
}
\left[(1+z^2)w_0 + z(1+z)w_1\right]
}{
2\sqrt{3}\,(1+z^2)\,\sqrt{\lambda}
},
\end{equation}
\begin{equation}\label{29}
q_{BA}(z) = -1 +
\frac{
(1+z)^{2+2w_0}
(1+z^2)^{-1+w_1}
\left(6\alpha H_0^2 + \beta\right)
\left[1+w_0
+ z\left(w_1 + z(1+w_0+w_1)\right)\right]
}{
-\beta
+
(1+z)^{2+2w_0}
(1+z^2)^{w_1}
\left(6\alpha H_0^2 + \beta\right)
},
\end{equation}
\begin{equation}\label{30}
\begin{aligned}
j_{BA}(z) &=
\frac{w_0+w_1}{(1+z^2)^2} +
\frac{1}{(1+z^2)^2}
\Bigg[
\frac{
\beta \left(-1 - z^2(2+z^2) + w_0 + w_1 \right)
}{-\beta + (1+z)^{2+2w_0}(1+z^2)^{w_1}(6H_0^2\alpha+\beta)} +
\frac{
(1+z)^{2+2w_0}(1+z^2)^{w_1}(6H_0^2\alpha+\beta)
}{-\beta + (1+z)^{2+2w_0}(1+z^2)^{w_1}(6H_0^2\alpha+\beta)}
\\
&\qquad \times
\Big(
w_0\!\left(2w_0 + z^2(2+z^2)(1+2w_0)\right) + z\!\left(4 + 2z + z^3 + 4(1+z)(1+z^2)w_0\right)w_1 + 2z^2(1+z)^2w_1^2
\Big)
\Bigg].
\end{aligned}
\end{equation}
\subsection{Jassal–Bagla–Padmanabhan (JBP) Parametrisation}

The JBP parametrisation employs a second-degree polynomial and is constructed to keep $ w(a) $ identical at the present epoch and at high redshift, while permitting variations at intermediate times \cite{Jassal:2004ej, Jassal:2005qc}. It is given by
\begin{equation}\label{31}
 w_{JBP}(a)=w_0 + w_1a(1-a),
 \qquad 
a = \frac{1}{1+z}.
\end{equation}
From Eqs. (\ref{16}), (\ref{18}), (\ref{31}) and $p=w\rho$, we get
\begin{equation}\label{32}
H_{JBP}(z)=\sqrt{
\frac{
\left(6\alpha H_{0}^{2}+\beta\right)
(1+z)^{2\left(1+w_{0}\right)}
\exp\!\left(\frac{w_{1} z^{2}}{(1+z)^{2}}\right)
-\beta
}{6\alpha}
}
\end{equation}
For JBP, the cosmological parameters are calculated as:
\begin{equation}\label{33}
\rho_{JBP}(z) =
\frac{
\sqrt{
e^{\frac{z^{2}w_{1}}{(1+z)^{2}}}
(1+z)^{2+2w_{0}}
\left(6H_{0}^{2}\alpha + \beta\right)
}
}{
2\sqrt{3}\,\sqrt{\lambda}
}
\end{equation}
\begin{equation}\label{34}
p_{JBP}(z) =
\frac{
\sqrt{
e^{\frac{z^{2}w_{1}}{(1+z)^{2}}}
(1+z)^{2+2w_{0}}
\left(6H_{0}^{2}\alpha + \beta\right)
}
\left[(1+z)^{2}w_{0} + zw_{1}\right]
}{
2\sqrt{3}\,(1+z)^{2}\sqrt{\lambda}
}
\end{equation}

\begin{equation}\label{35}
q_{JBP}(z) = -1 +
\frac{
e^{\frac{z^{2}w_{1}}{(1+z)^{2}}}
(1+z)^{2w_{0}}
\left(6H_{0}^{2}\alpha + \beta\right)
\left[(1+z)^{2}(1+w_{0}) + zw_{1}\right]
}{
-\beta +
e^{\frac{z^{2}w_{1}}{(1+z)^{2}}}
(1+z)^{2+2w_{0}}
\left(6H_{0}^{2}\alpha + \beta\right)
}
\end{equation}

\begin{equation}\label{36}
\begin{aligned}
j_{JBP}(z) =
\frac{1}{
-(1+z)^2 \beta
+ e^{\frac{z^2 \omega_1}{(1+z)^2}} (1+z)^{4+2\omega_0}
\left( 6 H_0^2 \alpha + \beta \right)
} \Bigg[ -(1+z)^2 \beta
+ 6 H_0^2 \alpha \, e^{\frac{z^2 \omega_1}{(1+z)^2}} (1+z)^{2\omega_0}
\Big[(1+z)^4 \omega_0 (1+2\omega_0)
\\[4pt]
 + (1+z)^2 (1+4z\omega_0)\omega_1
+ 2 z^2 \omega_1^2 \Big]
+ \beta \, e^{\frac{z^2 \omega_1}{(1+z)^2}} (1+z)^{2\omega_0}
\Big[(1+z)^4 \omega_0 (1+2\omega_0)
+ (1+z)^2 (1+4z\omega_0)\omega_1
+ 2 z^2 \omega_1^2 \Big]\Bigg].
\end{aligned}
\end{equation}

\subsection{Logarithmic (LOG) Parametrisation}
The logarithmic form ensures a smooth evolution of the equation of state parameter, making it suitable for cosmological data fitting and preventing abrupt behaviour \cite{Tripathi:2016slv}. It is defined as 
\begin{equation}\label{37}
w_{LOG}(a)=w_0-w_1lna ,
\qquad 
a = \frac{1}{1+z}.
\end{equation}
For LOG parameterisation, the Hubble parameter can be obtained as
\begin{equation}\label{38}
H_{LOG}(z)=\sqrt{
\frac{
\left(6\alpha H_{0}^{2}+\beta\right)
(1+z)^{2\left(1+w_{0}\right)}
\exp\!\left[w_{1}\left(\ln(1+z)\right)^{2}\right]
-\beta
}{6\alpha}
}
\end{equation}
From this, the corresponding cosmological parameters are reconstructed as:
\begin{equation}\label{39}
\rho_{LOG}(z) =
\frac{
\sqrt{
e^{w_{1}\,\ln(1+z)^{2}}
(1+z)^{2(1+w_{0})}
\left(6H_{0}^{2}\alpha + \beta\right)
}
}{
2\sqrt{3}\,\sqrt{\lambda}
}
\end{equation}

\begin{equation}\label{40}
p_{LOG}(z) =
\frac{
\sqrt{
e^{w_{1}\,\ln(1+z)^{2}}
(1+z)^{2+2w_{0}}
\left(6H_{0}^{2}\alpha + \beta\right)
}
\left(w_{0} + w_{1}\ln(1+z)\right)
}{
2\sqrt{3}\,\sqrt{\lambda}
}
\end{equation}

\begin{equation}\label{41}
q_{LOG}(z) = -1 +
\frac{
1 + w_{0} + w_{1}\ln(1+z)
}{
1 -
\displaystyle
\frac{
e^{-w_{1}\ln(1+z)^{2}}\,
(1+z)^{-2(1+w_{0})}\,\beta
}{
6H_{0}^{2}\alpha + \beta
}
}
\end{equation}
\begin{equation}\label{42}
\begin{aligned}
j_{LOG}(z) =\;&
w_{0}
+ 2w_{0}^{2}
+ w_{1}
+ w_{1}\ln(1+z)
+ 4w_{0}w_{1}\ln(1+z)
+ 2w_{1}^{2}\ln(1+z)^{2}
\\[4pt]
&+
\frac{
\beta \left[
-1 + w_{0}
+ 2w_{0}^{2}
+ w_{1}
+ w_{1}\ln(1+z)
\left(1 + 4w_{0}
+ 2w_{1}\ln(1+z)\right)
\right]
}{
-\beta
+ e^{w_{1}\ln(1+z)^{2}}
(1+z)^{2+2w_{0}}
\left(6H_{0}^{2}\alpha + \beta\right)
}.
\end{aligned}
\end{equation}

\section{Observational Data and Statistical Analysis}\label{data}

A reliable evaluation of extended gravity models requires testing their theoretical predictions with a diverse set of observational probes. In the present work, we constrain the free parameters of the $f(Q,\mathcal{L}_m)$ gravity framework using three independent and complementary cosmological datasets: Observational Hubble Data (OHD), the Dark Energy Survey five-year Type Ia supernova compilation (DES~5YR), and Baryon Acoustic Oscillation (BAO) measurements. Using these datasets together allows for a more robust reconstruction of the late–time expansion history of the Universe over a broad range of redshifts. The joint analysis also helps reduce degeneracies between the model parameters and leads to tighter constraints on the $f(Q,\mathcal{L}_m)$ gravity model.

The estimation of cosmological parameters is carried out through a Markov Chain Monte Carlo (MCMC) analysis implemented in Python using the emcee sampler. This method enables the multidimensional parameter space to be explored efficiently and identifies the regions favoured by observational data through likelihood maximisation. The resulting posterior probability distributions are examined at the $1\sigma$ and $2\sigma$ confidence levels. The correlated behaviour of the parameters is illustrated through one- and two-dimensional marginalised confidence contours, and The resulting best-fit parameter values are analysed along with their corresponding $1\sigma$ uncertainties.

For clarity, a summary of the observational datasets employed in this analysis, together with their redshift coverage and number of data points, is presented in Table \ref{tab:datasets}.

\begin{table}[h]
\centering
\caption{Overview of the observational datasets used in the present analysis.}
\setlength{\tabcolsep}{9pt}
\renewcommand{\arraystretch}{1.5}
\begin{tabular}{l l c c c}
\hline\hline
Observational probe & Survey / Source & Data points & Redshift range & References \\
\hline
Hubble expansion data & Cosmic Chronometers (OHD) & 77 & $0.07 - 2.36$ & \cite{Zhang:2012mp,Zhao:2017urm} \\
Type Ia supernovae & DES 5-year (DES 5YR) & 1829 & $0.01 - 1.20$ & \cite{DES:2018paw} \\
\hline
BAO measurements & 6dFGS & 1 & 0.106 & \cite{Beutler:2011hx}\\
 & SDSS MGS & 1 & 0.15 & \cite{Ross:2014qpa}\\
 & BOSS LOWZ & 1 & 0.32 & \cite{BOSS:2016wmc}\\
 & BOSS CMASS & 1 & 0.57 & \cite{BOSS:2016wmc}\\
 & WiggleZ & 3 & $0.44 - 0.73$ & \cite{Blake:2011en}\\
 & eBOSS / Ly-$\alpha$ & 14 & $0.38 - 2.40$ & \cite{BOSS:2017uab} \\
\hline
Total BAO data &  & 21 & $0.106 - 2.40$ & \\
Overall total &  & 1927 & $0.01 - 2.40$ &  \\
\hline\hline
\end{tabular}
\label{tab:datasets}
\end{table}

\subsection*{Observational Hubble Data (OHD)}

Observational Hubble Data offer direct estimates of the cosmic expansion rate over a range of redshifts. These measurements are commonly derived through the cosmic chronometer method, which uses the differential age evolution of passively evolving galaxies. Since this method does not require assumptions about a particular background cosmological model, it provides a largely model-independent way to investigate the late-time evolution of the Universe.

In the present study, we use a compilation of 77 Hubble parameter measurements covering the redshift interval $0.07 \le z \le 2.36$ \cite{Singh:2024kez, Stern:2009ep, BOSS:2017fdr, BOSS:2013rlg, BOSS:2016wmc, BOSS:2016zkm, Ratsimbazafy:2017vga, Moresco:2016mzx, Zhang:2012mp,Zhao:2017urm}. The Hubble parameter may be expressed in terms of the redshift as $H(z) = -\frac{1}{1+z}\frac{dz}{dt}$ where $z$ represents the cosmological redshift.

The contribution of the observational Hubble data to the likelihood analysis is evaluated through the chi-square estimator
\begin{equation}
\chi^2_{\rm H}
=
\sum_{i=1}^{77}
\frac{\left[H_{\rm th}(z_i)-H_{\rm obs}(z_i)\right]^2}
{\sigma_i^2},
\end{equation}
where $H_{\rm th}$ and $H_{\rm obs}$ denote the theoretical and observational values of the Hubble parameter, respectively, while $\sigma_i$ corresponds to the observational uncertainty associated with each measurement.

\subsection*{DES 5 Year Supernova Sample (DES 5YR)} 
Type Ia supernovae are among the most accurate observational tools for studying cosmic acceleration, primarily because they can be used as standardizable candles. Measurements of their luminosity–distance relation place strong constraints on the late-time expansion history of the Universe.
In the present study, we use the five-year Dark Energy Survey supernova sample (DES 5YR) \cite{DES:2024jxu, Colgain:2025fct, DES:2018paw}, consisting of 1829 spectroscopically confirmed events distributed over a broad redshift interval (see Table \ref{tab:datasets}). The DES 5YR dataset benefits from refined photometric calibration, consistent data processing, and careful consideration of observational uncertainties. These characteristics make it a suitable dataset for placing constraints on and evaluating extended gravity models.

The distance modulus is defined as
\begin{equation}
\mu_{\rm th} = 5\log_{10}(D_L) + \mu_0,
\end{equation}
with
\begin{equation}
\mu_0 = 5\log_{10}\left(\frac{c}{H_0}\right) + 25,
\end{equation}
where the luminosity distance is given by
\begin{equation}
D_L(z) = c(1+z)\int_0^z \frac{dz'}{H(z')}.
\end{equation}

The likelihood function is constructed using
\begin{equation}
\chi^2_{\rm S}
=
\Delta \mu_i \, C^{-1}_{ij} \, \Delta \mu_j ,
\end{equation}
where $\Delta \mu_i = \mu_{\rm th}(z_i)-\mu_{\rm obs}(z_i)$ and $C_{ij}$ is the full covariance matrix.

\subsection*{Baryon Acoustic Oscillation (BAO) Data}

Baryon Acoustic Oscillations originate from sound waves propagating in the photon-baryon plasma before recombination \cite{Peebles:1970ag}. The characteristic acoustic scale imprinted in the matter distribution acts as a standard ruler, enabling precise measurements of cosmological distances across different redshifts \cite{SDSS:2005xqv, Percival:2007yw}.

In the present analysis, BAO observations from several galaxy surveys including 6dFGS \cite{Beutler:2011hx}, SDSS MGS \cite{Ross:2014qpa}, BOSS LOWZ, BOSS CMASS \cite{BOSS:2016wmc}, WiggleZ \cite{Blake:2011en}, and eBOSS Ly-$\alpha$ \cite{BOSS:2017uab} are employed, yielding a total of 21 data points. These measurements provide strong geometric constraints that are largely insensitive to astrophysical systematics and have been extensively used in recent observational studies within modified gravity and dark energy scenarios \cite{Balhara:2023mgj, Singh:2024tur}.
 
The BAO observable is defined as
\begin{equation}
d_z = \frac{r_s(z_*)}{D_V(z)},
\end{equation}
with
\begin{equation}
D_V(z) =
\left[z\, d_A^2(z)\, H^{-1}(z)\right]^{1/3}.
\end{equation}

The corresponding chi-square estimator is written as
\begin{equation}
\chi^2_{\rm B}
=
\Delta d\, C^{-1}\, \Delta d^{T},
\end{equation}
where $\Delta d = d_z^{\rm obs} - d_z^{\rm th}$.

\subsection*{Joint Likelihood Analysis}

To obtain stronger and more robust constraints on the model parameters, the three observational datasets are analysed together in a combined statistical framework. The combined chi-square statistic is then expressed as
\begin{equation}
\chi^2_{\rm Joint}
=
\chi^2_{\rm H}
+
\chi^2_{\rm S}
+
\chi^2_{\rm B}.
\end{equation}

Here, $\chi^2_{\rm H}$ represents the contribution from the observational Hubble data, $\chi^2_{\rm S}$ corresponds to the likelihood from the DES~5YR supernova sample, and $\chi^2_{\rm B}$ denotes the contribution from baryon acoustic oscillation measurements. This combined approach efficiently exploits the complementary nature of the datasets and significantly reduces parameter degeneracies. In particular, the joint analysis improves constraints on the modified gravity parameters by simultaneously incorporating expansion-rate measurements, luminosity distances, and geometric distance indicators.

\begin{figure}\centering
	\includegraphics[scale=0.7]{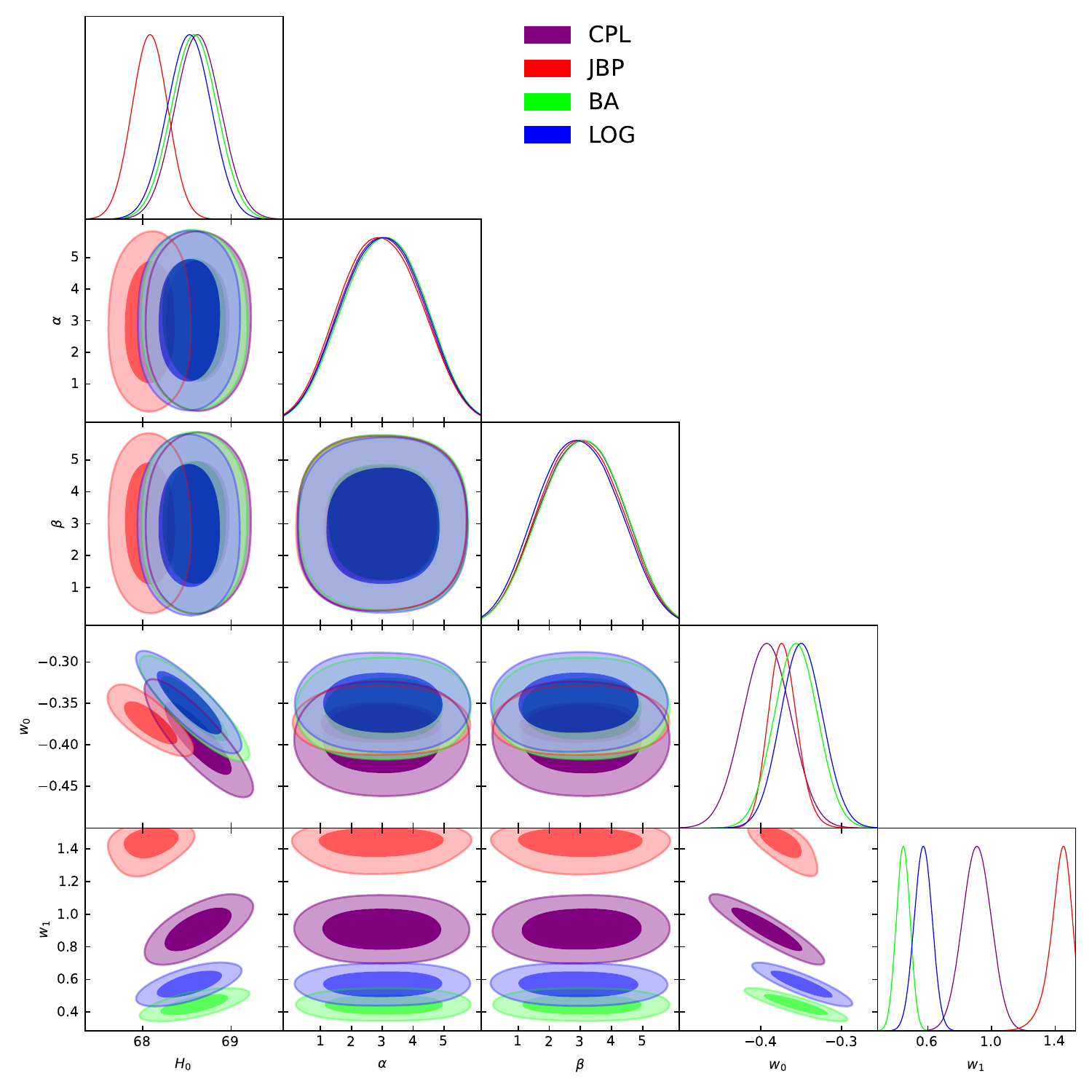}
\caption{ Posterior constraints on the $ f(Q, \mathcal{L}_m) $ model parameters at the $ 1\sigma $ and $ 2\sigma $ CLs.}
\label{fig:contours}
\end{figure}

Fig. \ref{fig:contours} presents the posterior distributions for the CPL, JBP, BA, and LOG parametrisations, including the corresponding one- and two-dimensional confidence regions at the $1\sigma$ and $2\sigma$ levels. Strong correlations are observed among $H_0$, $\alpha$, and $\beta$, reflecting the interplay between the background expansion and the gravitational sector, while the parameters
$w_0$ and $w_1$ remain comparatively stable across all parametrisations. The best-fit estimates of the model parameters, together with their associated $1\sigma$ uncertainties, are summarized in Table \ref{tab:bestfit}.

\begin{table*}[t]
\centering
\caption{Maximum likelihood estimates of the model parameters with $1\sigma$ CL for the CPL, JBP, BA, and LOG models for joint OHD+DES~5YR+BAO observations.}
\setlength{\tabcolsep}{14pt}
\renewcommand{\arraystretch}{1.8}
\begin{tabular}{lcccc}
\hline\hline
Parameter & CPL & JBP & BA & LOG \\
\hline
$H_0$ (KmS$^{-1}$Mpc$^{-1}$)
& $68.624^{+0.239}_{-0.239}$
& $68.079^{+0.186}_{-0.188}$
& $68.582^{+0.237}_{-0.239}$
& $68.525^{+0.231}_{-0.232}$ \\

$\alpha$
& $2.990^{+1.370}_{-1.369}$
& $2.943^{+1.382}_{-1.352}$
& $3.061^{+1.357}_{-1.368}$
& $3.023^{+1.363}_{-1.377}$ \\

$\beta$
& $3.040^{+1.354}_{-1.374}$
& $2.997^{+1.371}_{-1.350}$
& $3.057^{+1.368}_{-1.383}$
& $2.930^{+1.371}_{-1.359}$ \\

$w_0$
& $-0.393^{+0.028}_{-0.028}$
& $-0.373^{+0.016}_{-0.016}$
& $-0.356^{+0.025}_{-0.025}$
& $-0.349^{+0.024}_{-0.024}$ \\

$w_1$
& $0.908^{+0.084}_{-0.084}$
& $1.431^{+0.056}_{-0.060}$
& $0.445^{+0.041}_{-0.041}$
& $0.571^{+0.052}_{-0.052}$ \\
\hline\hline
\end{tabular}

\vspace{0.2cm}
\begin{minipage}{0.95\textwidth}
\footnotesize

\end{minipage}
\label{tab:bestfit}
\end{table*}




\section{The Cosmological Parameters}\label{param}
\begin{figure}[htbp]
    \centering
   \includegraphics[scale=0.58]{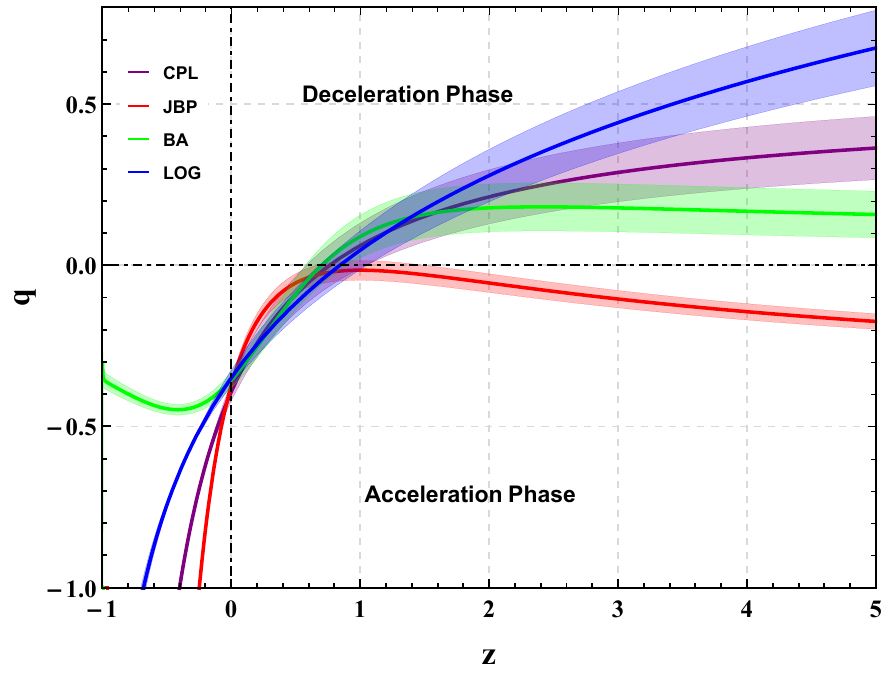}\hfill
\caption{FIG. 2: Redshift evolution of the deceleration parameter $q$.}
\label{fig:q}
\end{figure}

In this section, we examine the evolution of the principal kinematical and dynamical parameters reconstructed for the four EoS parametrisations, namely, CPL, JBP, BA, and LOG, to examine the physical consistency of the cosmological model. The Hubble parameter, which characterises the expansion rate of the Universe, is defined as
\begin{equation}
H=\frac{\dot a}{a}, \qquad 1+z=\frac{1}{a}.
\end{equation}
It plays a fundamental role in cosmology, since all other kinematical and dynamical quantities are constructed from the Hubble function and its derivatives with respect to redshift. The deceleration parameter, which determines whether the cosmic expansion is accelerating or decelerating, is given by
\begin{equation}
q=-\frac{\ddot a\,a}{\dot a^{2}}.
\end{equation}
The sign of the deceleration parameter $q$ dictates the nature of cosmic expansion; negative values of $q$ signify an accelerated expansion, while positive values of $q$ indicate a decelerating Universe. In Figure \ref{fig:q}, the reconstructed evolution of the deceleration parameter $q(z)$ demonstrates that the Universe undergoes a transition from an early decelerated phase to a late-time accelerated expansion for the CPL, BA, and LOG parametrisations, whereas the JBP model exhibits accelerated expansion throughout the entire cosmic evolution. For the CPL, BA, and LOG cases, the deceleration parameter remains positive at high redshifts, indicating a matter-dominated decelerating era. As the redshift decreases, $q(z)$ crosses zero and becomes negative in the recent epoch, signalling the onset of late-time cosmic acceleration \cite{Shaily:2024nmy}.



\begin{figure}[htbp]
    \centering
    \subfloat[]{\label{fig:rho}\includegraphics[scale=0.625]{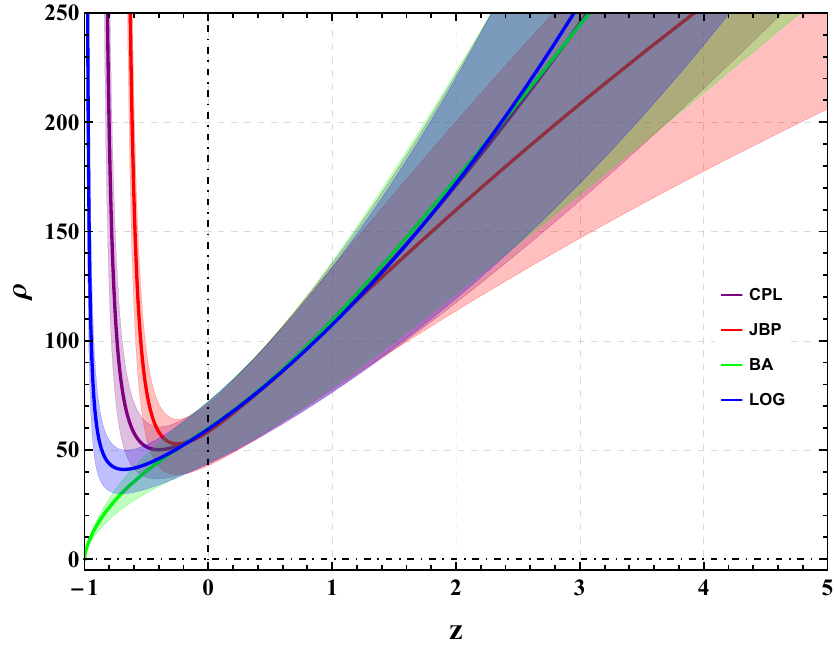}}\hfill
    \subfloat[]{\label{fig:p}\includegraphics[scale=0.64]{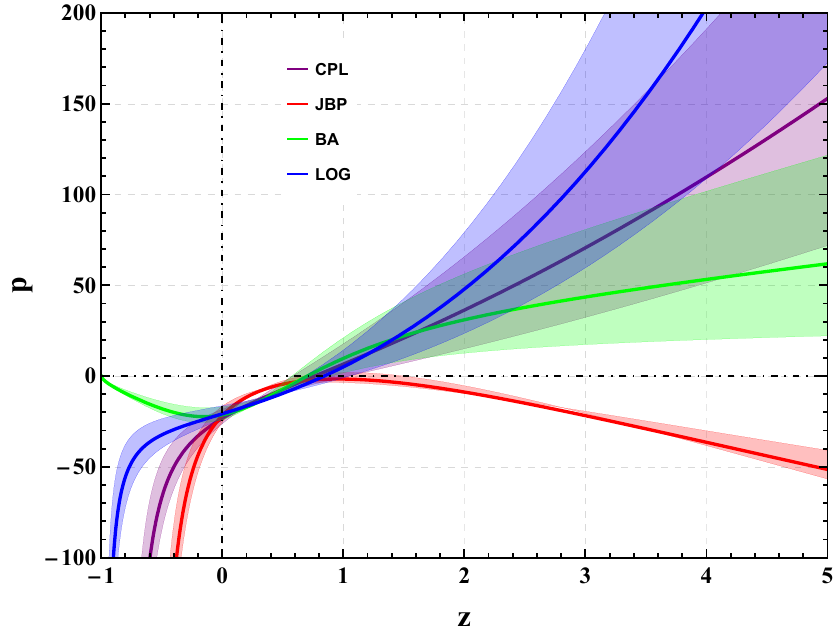}}\par
    \subfloat[]{\label{fig:w}\includegraphics[scale=0.65]{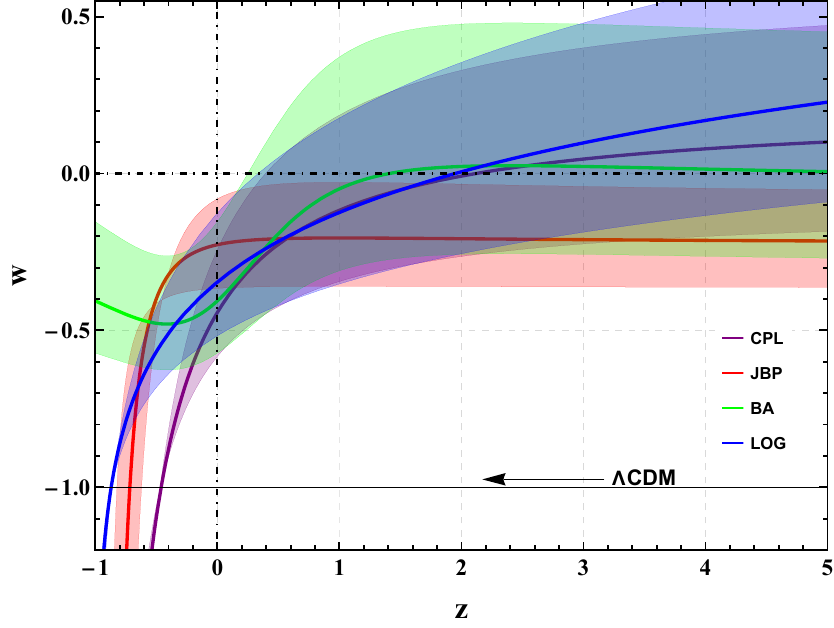}}
\caption{Evolution of the cosmological parameters: The energy density $\rho$, isotropic pressure $p$, and EoS parameter $ w $.}
    \label{fig:cosmological-parameters}
\end{figure}

Assuming a spatially flat FLRW universe, the cosmic dynamics in the present framework are governed by the modified Friedmann equations, in which the EoS parametrisations are encoded through effective energy density and pressure. As shown in Fig.~\ref{fig:rho}, the reconstructed profiles demonstrate that the effective energy density $\rho(z)$ remains strictly positive over the entire redshift range for all four parametrisations, ensuring the physical consistency of the modified gravitational framework. Its large values at high redshift reflect the dense early Universe, while its monotonic decrease with cosmic expansion indicates the gradual dilution of the effective cosmic fluid. The small differences among the CPL, JBP, BA, and LOG cases do not affect this overall trend, confirming the stability of the modified cosmological dynamics.

Within the same framework, the effective pressure $p(z)$ displayed in Fig. \ref{fig:p} plays a central role in determining the expansion history. At early times, $p(z)$ is predominantly positive, leading to attractive gravity and a decelerated expansion phase. As the Universe evolves, $p(z)$ decreases and becomes negative at low redshift for all parametrisations. In the modified Friedmann context, this transition enhances the repulsive contribution in the acceleration equation, providing the physical mechanism for the observed late-time accelerated expansion. The comparatively stronger negative pressure in the models indicates a more pronounced dark energy-driven modification of the cosmic expansion.
\begin{table}[ht]
\centering
\caption{Physical interpretation of the equation of state parameter $w$.}
\label{tab:eos_interpretation}
\renewcommand{\arraystretch}{1.35}

\begin{tabular}{c l l l}
\hline\hline
\label{w}
\textbf{$w$ range} & \textbf{Cosmic Phase}  &  \textbf{Physical Behaviour} & \textbf{Status of the Universe} \\
\hline

$ w > 1 $ 
& Super stiff era, Ekpyrotic universe 
& Extremely high pressure 
& Pre-Big Bang phases \\

$w = 1$ 
& Stiff--fluid phase 
& Kinetic energy dominated 
& Early time only \\

$\tfrac{1}{3} < w < 1$ 
& Ultra--relativistic phase 
& Strong radiation pressure 
& Outside late time regime \\

$w = \tfrac{1}{3}$ 
& Radiation era 
& Relativistic matter 
& Early cosmic phase \\

$0 < w < \tfrac{1}{3}$ 
& Warm--matter phase 
& Intermediate behaviour 
& Pre dark energy stage \\

$w = 0$ 
& Matter era 
& Pressureless dust 
& High--redshift regime \\

$-\tfrac{1}{3} < w < 0$ 
& Transition regime 
& Weak negative pressure 
& Onset of acceleration \\

$w = -\tfrac{1}{3}$ 
& Acceleration boundary 
& Zero deceleration 
& Transition point \\

$-1 < w < -\tfrac{1}{3}$ 
& Quintessence regime 
& Dynamical dark energy 
& Main late-time regime \\

$w = -1$ 
& de Sitter phase 
& Vacuum energy expansion 
& $\Lambda$CDM limit \\

$w < -1$ 
& Phantom regime 
& Superacceleration 
& Not supported \\

\hline\hline
\end{tabular}
\end{table}
The equation-of-state parameter $w(z)$, reconstructed using the CPL, JBP, BA, and LOG equation-of-state parametrisations, serves as a crucial probe of the dynamical behaviour of dark energy. As shown in Fig. \ref{fig:w}, despite their different functional forms, the models exhibit a broadly consistent evolutionary pattern. The physical interpretation of the different ranges of $w(z)$ is summarized in Table~\ref{w}. At early times, the CPL, BA, and LOG parametrisations yield positive values of $w$, indicating matter-dominated expansion, and then transition to negative values near the present epoch. At present, all parametrisations yield negative values of the equation of state parameter, placing the cosmic fluid within the quintessence regime, which successfully accounts for the observed accelerated expansion of the Universe. At late times, the CPL, JBP, and LOG parametrisations enter the phantom region, signalling an unstable phase of cosmic acceleration, whereas the BA model maintains a stable quintessence-like behaviour.

\begin{figure}[htbp]
    \centering
    \subfloat[]{\label{fig:c}\includegraphics[scale=0.64]{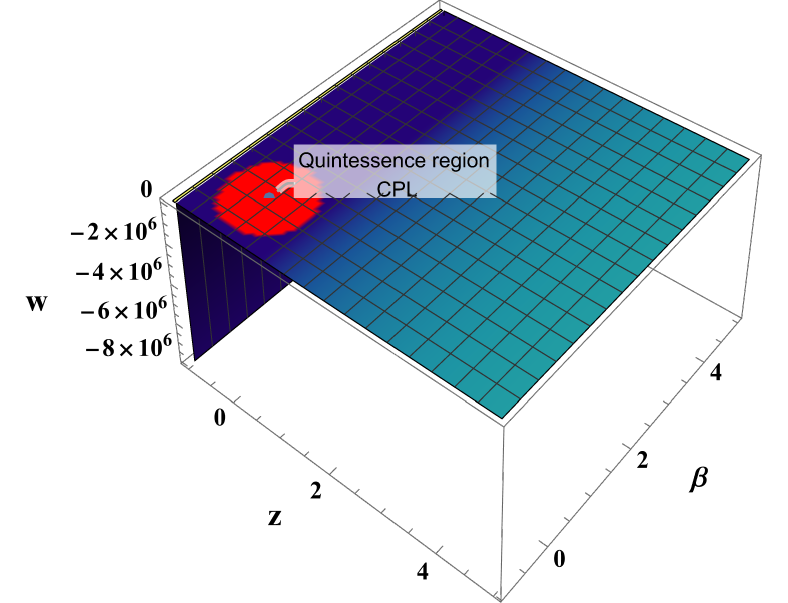}}\hfill
    \subfloat[]{\label{fig:j}\includegraphics[scale=0.64]{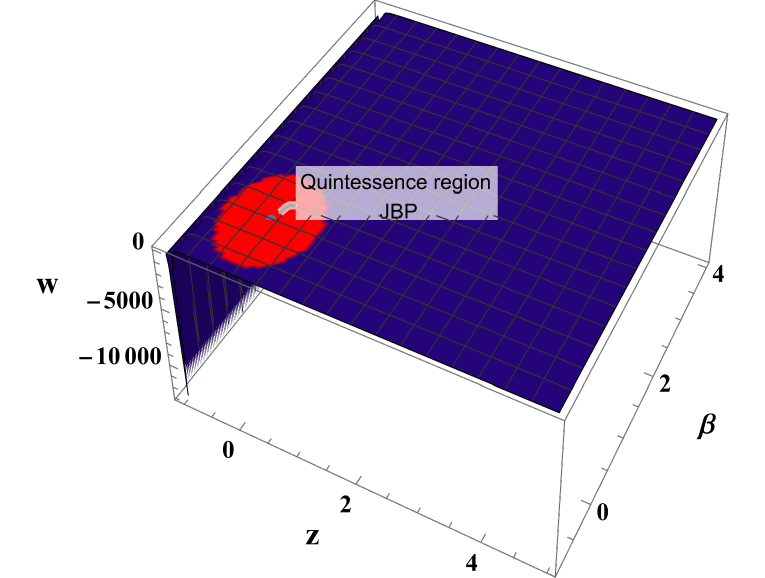}}\par
    \subfloat[]{\label{fig:b}\includegraphics[scale=0.64]{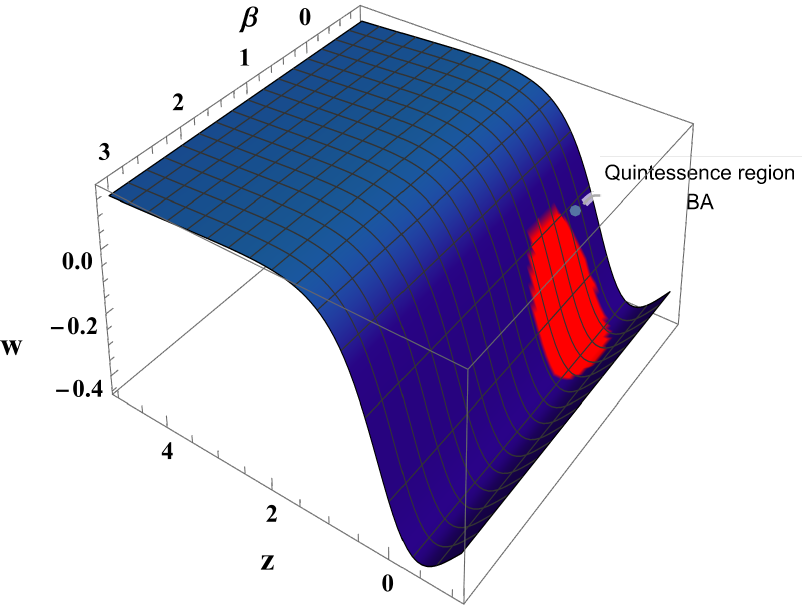}}\hfill
     \subfloat[]{\label{fig:l}\includegraphics[scale=0.64]{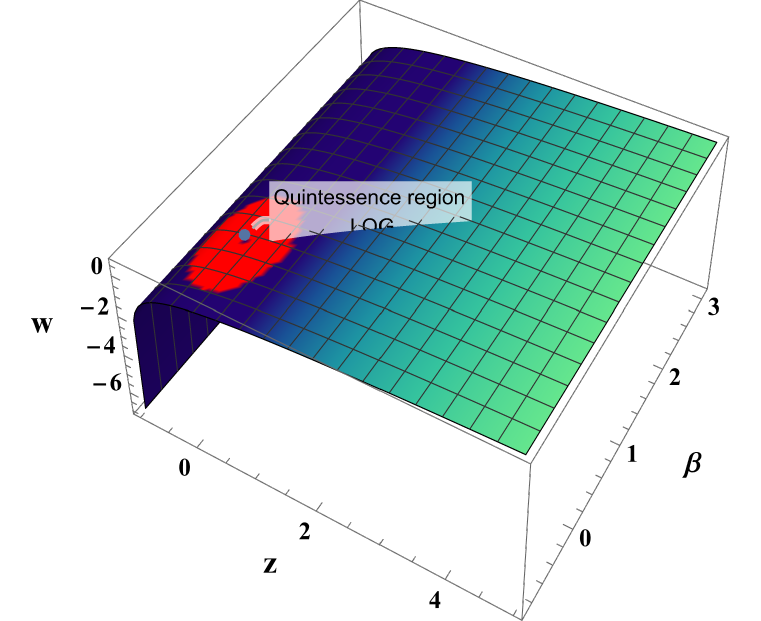}}
\caption{The embedded quintessence region over the CPL, JBP, BA, and LOG EoS parametrisations for OHD+DES~5YR+BAO observations, respectively.}
    \label{fig:EOS-3d}
\end{figure}

In dynamical dark energy models, the EoS $ w $ varies with redshift. In Fig.~\ref{fig:EOS-3d}, we see that the quintessence region $ (-1/3 > w > -1) $ is embedded in the model using the CPL, JBP, BA, and LOG equation-of-state parametrisations for the OHD+DES~5YR+BAO dataset at late times \cite{Shaily:2026xdk}.

\section{Credibility of the Energy Conditions}\label{ec}

Energy conditions (ECs) provide an important framework for examining the physical properties of spacetime and the behaviour of null (lightlike), timelike, and spacelike geodesics in gravitational theories. These conditions impose basic restrictions on the stress-energy tensor $T_{ij}$ through constraints on the energy density $\rho$ and pressure $p$, and therefore play a significant role in the study of spacetime singularities, stability, and cosmological evolution.

Their investigation becomes particularly relevant in modified gravity theories, where effective cosmic fluids may emerge as a consequence of nonlinear matter-geometry couplings. In such frameworks, energy conditions offer a useful criterion for assessing the physical viability of the reconstructed cosmological scenarios. These conditions can be motivated by the well-known Raychaudhuri equations \cite{Raychaudhuri:1953yv, Kar:2006ms}, which describe the evolution of geodesic congruences and link spacetime focusing directly to the matter content of the Universe. For a timelike geodesic congruence, the Raychaudhuri equation can be expressed as
\begin{equation}
\frac{d\theta}{d\tau}
= -\frac{1}{3}\theta^{2}
- \sigma_{\mu\nu}\sigma^{\mu\nu}
+ \omega_{\mu\nu}\omega^{\mu\nu}
- R_{\mu\nu}u^\mu u^\nu ,
\end{equation}
Here $\theta$ represents the expansion scalar, while $\sigma_{\mu\nu}$ and $\omega_{\mu\nu}$ correspond to the shear and rotation tensors, respectively, and $R_{\mu\nu}$ denotes the Ricci tensor. For null geodesics characterized by the vector field $k^\mu$, the analogous relation is given by
\begin{equation}
\frac{d\theta}{d\tau}
= -\frac{1}{2}\theta^{2}
- \sigma_{\mu\nu}\sigma^{\mu\nu}
+ \omega_{\mu\nu}\omega^{\mu\nu}
- R_{\mu\nu}k^\mu k^\nu .
\end{equation}

Since the shear term satisfies $\sigma_{\mu\nu}\sigma^{\mu\nu}\ge0$ and for hypersurface orthogonal congruences the rotation vanishes $(\omega_{\mu\nu}=0)$, the condition for gravity to remain attractive $(d\theta/d\tau<0)$ leads to the inequalities
\begin{equation}
R_{\mu\nu}u^\mu u^\nu \ge 0 \quad (\text{SEC}),
\end{equation}
\begin{equation}
R_{\mu\nu}k^\mu k^\nu \ge 0 \quad (\text{NEC}) .
\end{equation}

In the $f(Q, \mathcal{L}_m)$ framework, these geometric relations are translated into physical conditions by expressing the curvature terms through the effective energy-momentum tensor. This yields
\begin{equation}
\left(T_{\mu\nu}^{\rm eff}-\frac{1}{2}T^{\rm eff}g_{\mu\nu}\right)u^\mu u^\nu \ge 0,
\qquad
T_{\mu\nu}^{\rm eff}k^\mu k^\nu \ge 0 .
\end{equation}

For a perfect fluid description of the effective cosmic medium,
\begin{equation}
T_{\mu\nu}^{\rm eff}=(\rho_{\rm eff}+p_{\rm eff})u_\mu u_\nu - p_{\rm eff}g_{\mu\nu}.
\end{equation}
These relations reduce to the familiar algebraic forms
\begin{equation}
\rho_{\rm eff}+3p_{\rm eff}\ge0 \quad (\text{SEC}), 
\qquad 
\rho_{\rm eff}+p_{\rm eff}\ge0 \quad (\text{NEC}) .
\end{equation}
\begin{figure}\centering
	\subfloat[]{\label{anec}\includegraphics[scale=0.6]{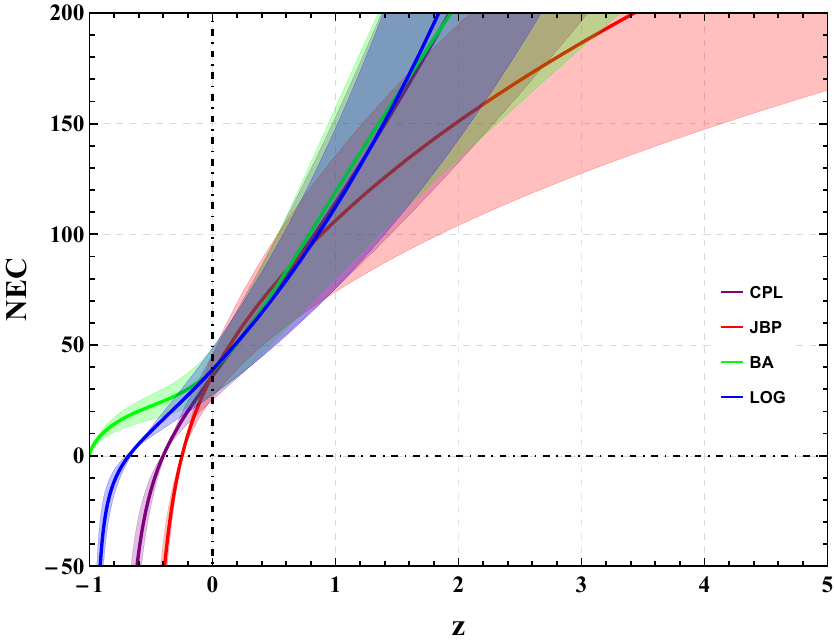}}\hfill
	\subfloat[]{\label{cdec}\includegraphics[scale=0.6]{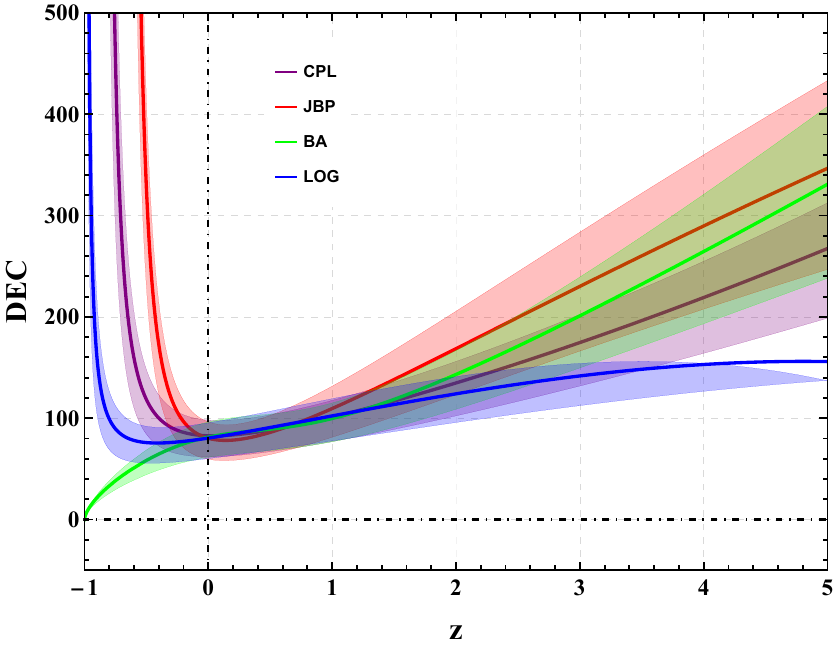}}\par 
	\subfloat[]{\label{bsec}\includegraphics[scale=0.6]{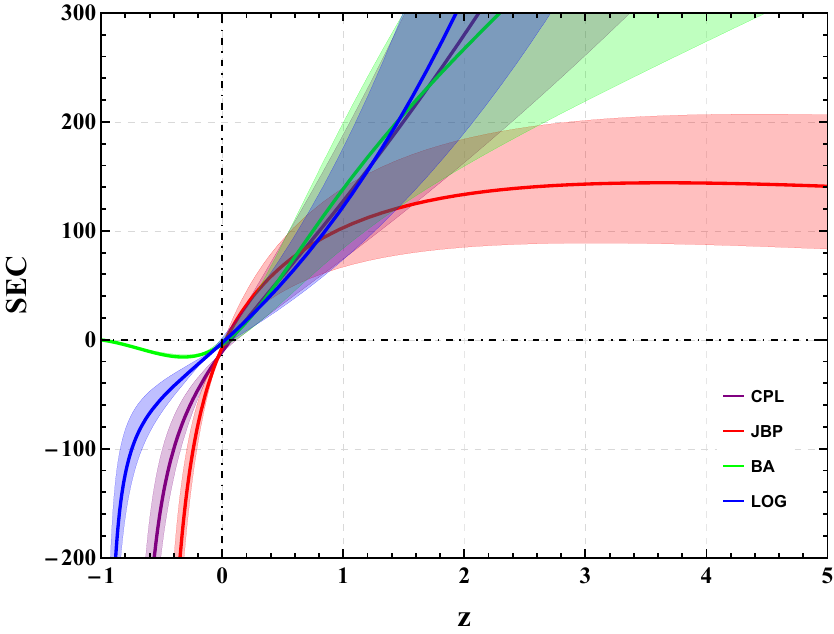}}
	\caption{ The variations of the energy conditions for OHD+DES~5YR+BAO observations.}
	\label{ecs}
\end{figure}
Thus, even in $f(Q, \mathcal{L}_m)$ gravity, the Raychaudhuri equation remains the fundamental theoretical basis for defining the null and strong energy conditions, although these conditions now constrain the effective cosmic fluid generated by the nonlinear matter geometry coupling rather than ordinary matter alone. In the present work, we focus on the null, strong, and dominant energy conditions for assessing the physical viability of late-time cosmological models. For a perfect fluid matter distribution, the energy conditions may be expressed through the effective energy density $\rho_{eff}$ and pressure $p_{eff}$ as:\

\begin{itemize}
  \item \textbf{SEC:} ${\rho_{eff}}+3{p_{eff}} \ge 0 \;\Rightarrow\; \rho_*\footnote{* Denotes the value of the parameter for the chosen EoS, namely CPL, JBP, BA, and LOG.}+3p_* \ge 0$,
  \item \textbf{NEC:} ${\rho_{eff}}+{p_{eff}} \ge 0 \;\Rightarrow\; \rho_*+p_* \ge 0$,
  \item \textbf{DEC:} ${\rho_{eff}} \ge 0,\; |{p_{eff}}| \le {\rho_{eff}} 
  \;\Rightarrow\; \rho_* \ge 0,\; |p_*| \le \rho_*$.
\end{itemize}
These results agree with earlier studies in modified gravity, including the analysis by Capozziello et al. \cite{Capozziello:2006dj}.

\subsection*{Energy Conditions in the Present Model}

To assess the physical viability of the reconstructed cosmological solutions, we analyse the behaviour of the null, dominant, and strong energy conditions using the effective energy density $\rho(z)$ and pressure $p(z)$. Fig. \ ref {ecs} shows the redshift evolution of the combinations $\rho_*+p_*$, $\rho_*-p_*$, and $\rho_*+3p_*$ corresponding to the NEC, DEC, and SEC, respectively, for the CPL, BA, JBP, and LOG parametrisations.
\begin{table}[htbp]
\centering
\caption{Energy conditions and dynamical behaviour of dark energy models at present and future epochs.}
\label{tab:dynamical_behaviour}

\setlength{\tabcolsep}{18pt}
\renewcommand{\arraystretch}{1.5}

\begin{tabular}{lccccc c}
\hline\hline
Model & Epoch & Redshift & NEC & DEC & SEC  \\
\hline
CPL & Present & $z=0$ & $\checkmark$\footnote{$\checkmark$ denotes that the corresponding energy condition is satisfied } & $\checkmark$ & $\times$\footnote{$\times$ denotes violation at the given epoch}  \\
    & Future  & $z\to -1$ & $\checkmark$ & $\checkmark$ & $\times$ \\

JBP & Present & $z=0$    & $\checkmark$ & $\checkmark$ & $\times$ \\
    & Future  & $z\to -1$ & $\checkmark$ & $\checkmark$ & $\times$ \\

BA  & Present & $z=0$    & $\checkmark$ & $\checkmark$ & $\times$ \\
    & Future  & $z\to -1$ & $\checkmark$ & $\checkmark$ & $\times$  \\

LOG & Present & $z=0$    & $\checkmark$ & $\checkmark$ & $\times$ &  \\
    & Future  & $z\to -1$ & $\checkmark$ & $\checkmark$ & $\times$ \\
\hline\hline
\end{tabular}
\vspace{2mm}
\end{table}

From figure \ref{anec}, it is observed that the null energy condition remains satisfied for all parametrisations over the entire redshift range in the case of observation BA, which indicates that the effective cosmic fluid does not exhibit exotic behaviour and that the causal structure of spacetime is preserved. In the case of other observations, the violation of the NEC at late times, where the sum of total energy density and pressure $ \rho_{\text{tot}} + P_{\text{tot}} $ drops below zero, can induce the accelerated cosmic expansion with a dark energy equation-of-state parameter $ w_{\text{DE}} < -1 $ (phantom dark energy), though recent analyses suggest that under standard total-fluid best-fit cosmological data models, no concrete macroscopic NEC violation is strictly observed.

In Fig. \ref{cdec}, the dominant energy condition stays positive throughout the cosmic evolution, ensuring that the energy density consistently dominates over pressure and that energy propagation remains causal. Also, it leads to several fundamental conclusions about the physical nature of the universe's matter, energy flow, and long-term fate of the universe as follows: \\
(i) Mass-energy can never be observed flowing faster than the speed of light. Energy propagation remains strictly bounded by local causality, \\
(ii) Protected against a Big Rip singularity, a catastrophic scenario where an exponentially growing dark energy density tears apart galaxies, atoms, and spacetime itself,\\
(iii) The cosmic vacuum remains stable. If the DEC were violated such that regions of negative energy density could manifest classically, the empty vacuum could spontaneously destabilise, splitting into unbounded regions of positive and negative energy,\\
(iv) It precludes the physical macro-existence of exotic phenomena like Traversable wormholes, etc.

Figure \ref{bsec} shows the violation of the strong energy condition at late times. In the neighbourhood of the present epoch, $\rho_*+3p_*$ becomes negative for all models, signalling the transition from the matter-dominated era to the dark energy-dominated era during the cosmic evolution which is consistent with earlier investigations in modified gravity \cite{Capozziello:2006dj, Visser:1999de, Moraes:2017zgm, Mandal:2020lyq}, and confirm that the reconstructed cosmological solutions in $f(Q, \mathcal{L}_m)$ gravity satisfy the fundamental requirements of causal consistency and physical viability while naturally accounting for the observed late-time accelerated expansion of the Universe. These results are summarised in Table \ref{tab:dynamical_behaviour}.

\section{ Stability analysis}\label{sa}

\qquad The stability of the $f(Q, \mathcal{L}_m)$ cosmological model may be examined by considering the propagation of perturbations through the cosmic fluid.
This quantity, commonly described by the sound speed, determines the rate at which density perturbations propagate through the cosmological medium. In a universe filled with a perfect fluid, the sound speed squared is fundamentally related to the pressure and energy density variations, defined as \cite{Peebles:2002gy, Rani:2026uxv}:
\begin{equation}\label{28}
C_s^2=\frac{dp}{d\rho}.
\end{equation}
For the model to be physically acceptable, thermodynamic stability requires the squared sound speed to satisfy $0 \leq C_s^2 \leq 1$, ensuring stable propagation of perturbations with subluminal velocity.

The square of the speed of sound $ C_s^2 $ is defined as $ \frac{dp}{d\rho} $, where the subscript $ s $ denotes constant entropy, and is stated in units of the speed of light. 

\begin{figure}\centering
\subfloat[]{\label{cs} \includegraphics[scale=0.65]{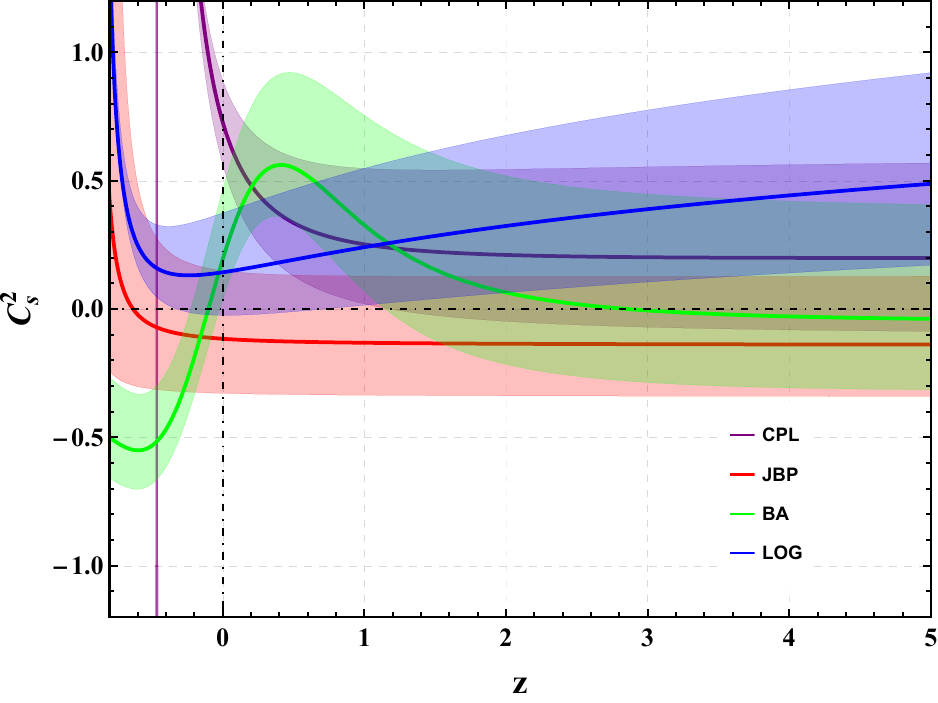}}\hfill
\caption{Evolution of stability parameters for OHD+DES~5YR+BAO observations.}
\label{vos}
\end{figure}

In this analysis, we compute the squared sound speed $ C_s^2 $ with the evolutionary profile displayed in Fig. \ref{vos}, revealing significant stability well within the range $ 0 \leq C_s^2 \leq 1 $ for all the models for OHD+DES~5YR+BAO observations.

\section{Concluding Remarks}\label{conclusion}
In this work, we have studied the late-time cosmological evolution of the Universe within the framework of $f(Q,\mathcal{L}_m)$ gravity using the model $f(Q, \mathcal{L}_m)=-\alpha Q+\lambda (2\mathcal{L}_m )^2+\beta$. This model introduces a nonlinear coupling between the non-metricity scalar and the matter Lagrangian, allowing the gravitational sector to influence the cosmic dynamics through matter–geometry interaction. Such a coupling modifies the effective cosmological behaviour and can provide a geometric mechanism for explaining the accelerated expansion of the Universe. To examine the model's implications, we reconstructed the Universe's cosmic expansion history using the CPL, JBP, BA, and logarithmic parametrisations of the dark-energy equation of state, which allow the EoS parameter to evolve with redshift and describe the model's possible dynamical behaviour. The model parameters were constrained through a combined analysis incorporating Observational Hubble Data (77 points), DES 5YR (1829 points), and Baryon Acoustic Oscillation (21 points) measurements. The obtained constraints indicate that the considered $ f(Q,\mathcal{L}_m) $ model remains compatible with the current observational description of the expansion history of the Universe. The posterior distribution of the model at the $ 1\sigma $ and $ 2\sigma $ CLs can be seen in Fig. \ref{fig:contours}, and the present values of the model parameters are given in Table \ref{tab:bestfit}.

\label{contours}
\begin{figure}\centering
	\includegraphics[width=1\linewidth, height=0.55\linewidth]{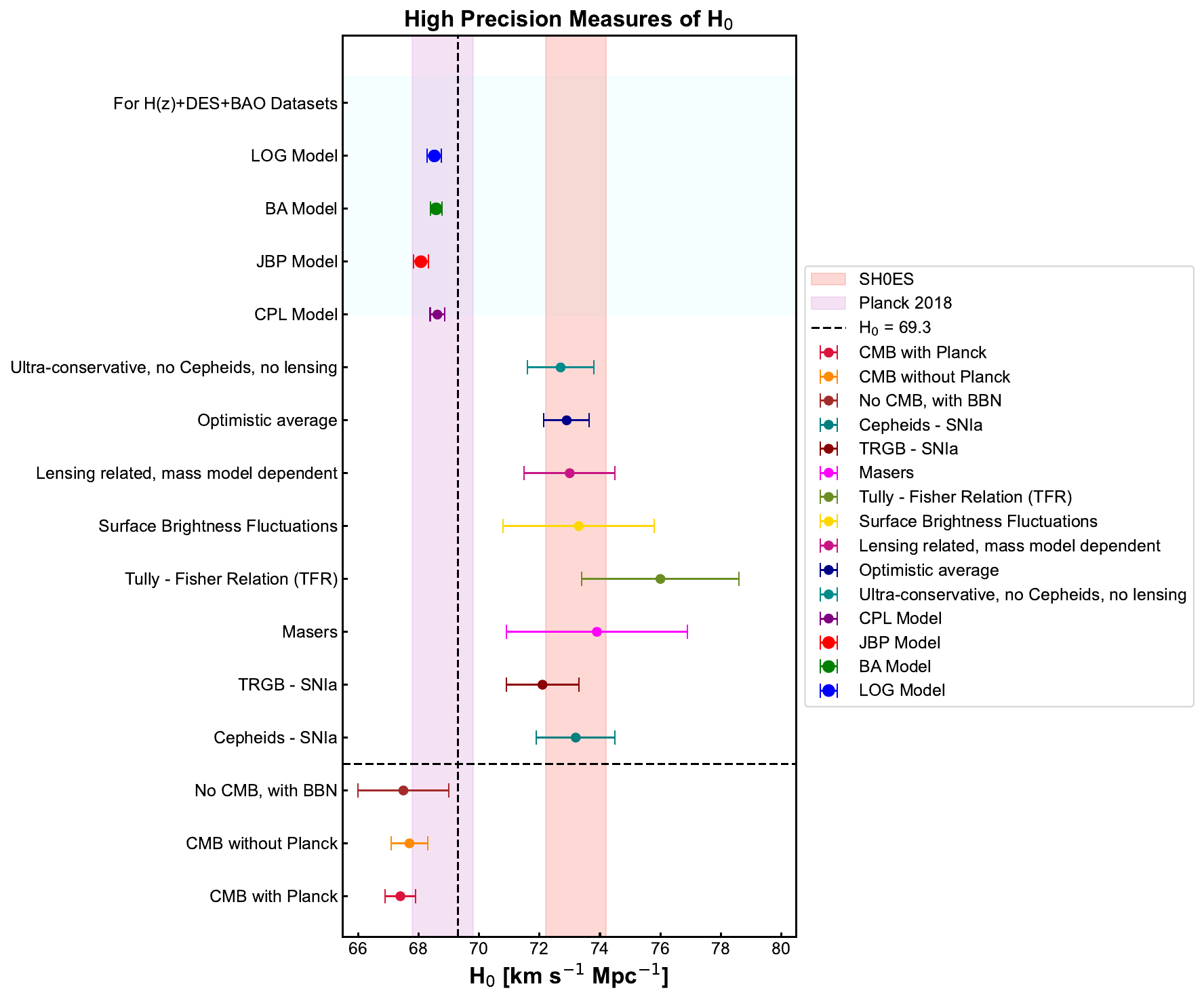}
		\caption{ \it The recent approximations of the $ H_0 $-tension evaluated from different observations.}
		\label{whisk}
\end{figure}

In figure \ref{fig:q}, the deceleration parameter $ q $ evolves from a decelerating phase at earlier times to an accelerated phase at the present epoch. This indicates that the cosmic expansion history cannot be fully described by a constant cosmological term alone and instead favours a mildly dynamical dark energy-like behaviour emerging from the modified gravitational sector. At the present epoch, the deceleration parameter takes the following values: $-0.393$, $-0.373$, $-0.355$, and $-0.349$ for the CPL, JBP, BA, and LOG models, respectively, indicating that all models exhibit an accelerated expanding Universe. 

The effective energy density stays positive during the entire cosmic evolution, supporting the physical consistency of the model. The effective pressure becomes negative at low redshift, providing the repulsive contribution required to sustain accelerated expansion. These features show that the model offers a coherent geometric mechanism for late-time acceleration, as illustrated in Fig. \ref{fig:cosmological-parameters}. The trajectories of the equation-of-state parameter in Fig. \ref{fig:w} undergo different stages from the early universe to the late universe. At the present epoch, all the parametrisations exist in the quintessence regime $-1<w_0<-\tfrac{1}{3}$. At late times, the CPL, JBP, and LOG parametrisations tend to the phantom region, indicating an unstable state of the Universe except in the case of the BA model, which remains in the quintessence regime.

In Figure \ref{fig:EOS-3d}, we have noticed that the embedded quintessence region $-1<w_0<-\tfrac{1}{3}$ is observed over the CPL, JBP, BA, and LOG EoS parametrisations for the OHD+DES~5YR+BAO observations, which is in good agreement with the observational limit of the $ w $CDM. 

The standard energy conditions are presented in Fig. \ref{ecs} to assess the physical consistency of the model. All energy conditions are satisfied at the present epoch, whereas the null and strong energy conditions are violated at late times. The violation of the NEC at late times, where the sum of total energy density and pressure $ \rho_{\text{tot}} + p_{\text{tot}} $ drops below zero, can induce the accelerated cosmic expansion with a dark energy equation-of-state parameter $ w_{\text{DE}} < -1 $ (phantom dark energy), though recent analyses suggest that under standard total-fluid best-fit cosmological data models, no concrete macroscopic NEC violation is strictly observed. 

In Figure \ref{bsec}, the late-time violation of the strong energy condition reflects the transition of the Universe from a matter-dominated phase to a dark energy-dominated phase during its cosmic evolution. This is consistent with earlier investigations in modified gravity and confirms that the reconstructed cosmological solutions in $f(Q, \mathcal{L}_m)$ gravity, which satisfy the fundamental requirements of causal consistency and physical viability while naturally accounting for the observed late-time accelerated expansion of the Universe. In Fig. \ref{vos},  we compute the squared sound speed $ C_s^2 $, with the evolutionary profile revealing significant stability well within the range $ 0 \leq C_s^2 \leq 1 $ for all the models for OHD+DES~5YR+BAO observations.

In conclusion, our results show that our model provides a viable geometric explanation for the late-time acceleration of the Universe while remaining consistent with current observational constraints. The study also highlights the importance of considering different EoS parametrisations when investigating modified gravity cosmologies. The recent assessments of the $ H_0 $-tension calculated from various observations are shown in Fig. \ref{whisk}. In our future work, we will try to extend the present analysis to include cosmological perturbations, structure formation, and gravitational wave constraints, which could provide more insights into the viability of matter–geometry coupling theories as alternatives to the standard dark energy paradigm.

\vskip0.2in
\section*{Acknowledgement} The authors thank Bennett University, Greater Noida, India, for providing the necessary facilities where a part of this work has been completed. 

\vskip0.2in

\textbf{\noindent Data Availability Statement:} In this manuscript, we used the observational data available in the literature. Our work does not produce any new data.

\vspace{0.4cm}

\textbf{\noindent Conflict of Interest:} The authors declare that they have no conflict of interest related to this work.

\end{document}